\documentstyle{article}
\newfont{\bbb}{msym10}
\newfont{\brm}{cmss10 scaled 1000}
\def\ra{\rightarrow}

\def\Z{\mbox{\bbb Z}}
\def\N{\mbox{\bbb N}}
\def\A{\mbox{${\cal A}$}}
\def\ph{\varphi}
\def\blacksquare{\quad \vrule height 6pt width 6pt}
\newenvironment{pf}
{{\bf\underline{proof}}:}{{\samepage\hfill\hbox{\blacksquare}}\par}
\newenvironment{thm}{\hfill\\\em}{\rm}
\newenvironment{lma}{\hfill\\\em}{\rm}

\def\ie{\hbox{\it i.e. }}
\def\ol#1{\overline{#1}}
\begin{document}
\baselineskip=20pt
\title{On Some Algebraic Structures Arising in String Theory}
\author{
Michael Penkava\\
University of California\\
Davis, CA 95616\\
michae@math.ucdavis.edu
\and Albert Schwarz\\
University of California\\
Davis, CA 95616\\
asschwarz@math.ucdavis.edu
\date{}
}
\maketitle
\def\he{\hat\epsilon}
\def\e{\epsilon}
\begin{abstract}
Lian and Zuckerman proved that the homology of a topological chiral
algebra can be
equipped with the structure of a BV-algebra; \ie one can introduce a
multiplication, an odd bracket, and an odd operator $\Delta$ having the
same properties as the corresponding operations in Batalin-Vilkovisky
quantization procedure.
We give a simple proof of their
results and discuss a generalization of these results to the non chiral
case. To simplify our proofs we use the following theorem giving a
characterization of a BV-algebra in terms of multiplication and an
operator $\Delta$:
{\em If $\A$ is a supercommutative, associative algebra and $\Delta$ is
an odd second order derivation on $\A$ satisfying $\Delta^2=0$, one can
provide $\A$ with the structure of a BV-algebra.}
\end{abstract}
\def\lbr#1#2{\left\lbrack#1,#2\right\rbrack}
The concept of a super commutative, associative
 algebra equipped with an additional
anti-bracket (odd bracket)
structure $\{\cdot,\cdot\}$ which satisfies the following
relations:
\begin{equation}
\{g,f\}=(-1)^{(f+1)(g+1)+1}\{f,g\} \quad\mbox{(super anticommutativity)}
\label{ga}
\end{equation}
\begin{equation}
\{f\{g,h\}\}=\{\{f,g\}h\}+(-1)^{(f+1)(g+1)}\{g\{f,h\}\}
\quad\mbox{(super Jacobi identity)}\label{ji}
\end{equation}
as well as
\begin{equation}
\{f,gh\}=\{f,g\}h +(-1)^{(f+1)g}g\{f,h\}
\quad\mbox{(super derivation rule)}\label{ld}
\end{equation}
was first seen in Gerstenhaber's 1963 article on cohomology of
associative rings \cite{ger}. (Here we use the convention that in superscripts
$f$ stands for the parity of $f$, where $f$ is assumed to be homogeneous.)
By an odd bracket, we mean that the parity of $\{f,g\}$ is opposite to
the parity of $fg$.
Note that if we consider the grading with the reversed parity, the
first two properties yield the structure of a super Lie algebra,
but the
multiplication is an additional structure on the algebra, so we shall
call an algebra satisfying relations
(1), (2) and (3) a G-algebra (Gerstenhaber
algebra). Later such algebras began appearing in different problems in
physics and mathematics, in particular, in the Batalin-Vilkovisky
approach to quantization of gauge theories\cite{bv,wit,sch,sc2}.
In \cite{zuk}, it was shown
that for every topological chiral conformal algebra, where all elements have
integral weights (conformal dimensions), the corresponding homology
naturally has the structure of a G-algebra.
In these two
latter cases, the G-algebra is also equipped with an odd operator $\Delta$,
which satisfies
\begin{equation}
\Delta(fg)=(\Delta f)g + (-1)^ff\Delta g + (-1)^f\{f,g\}\label{wb}
\end{equation}
and
$\Delta^2=0$. We will call a G-algebra equipped with such an  operator
$\Delta$ a BV-algebra (Batalin-Vilkovisky algebra). The term used for
such an algebra in \cite{zuk} is coboundary Gerstenhaber algebra or CGA.

One of the goals of the present paper is to simplify the constructions
and proofs of \cite{zuk}.  These constructions are closely related to
the papers \cite{wit2} and \cite{ver} that are devoted formally to the strings
in the $c=1$ background, but actually contain some general
considerations.
We consider also the case of a topological
conformal algebra containing the two Virasoro subalgebras (left and
right Virasoro algebras) without the assumption that all weights are integral.
We prove that in this case the corresponding BV-algebra also can be
constructed and discuss briefly the relation between this construction
and string theory.

To prove that our constructions really lead to BV-algebras we use a
theorem essentially giving a description of BV-algebras in terms of
multiplication and the operator $\Delta$. (The expressability of
$\{f,g\}$ in these terms follows immediately from equation (\ref{wb});
see \cite{wit}.)  We will begin with the formulation and proof of the
theorem, that permits us to simplify the verification of the axioms of
a BV-algebra.

First of all we recall that an operator $\alpha$ acting on a
$\Z_2$-graded algebra $\A$ is called a (super) derivation if for
any two elements $a,b\in \A$, we have
\begin{equation}
\lbr\alpha{\hat a}b=\alpha(a)b\label{sbr}
\end{equation}
Here $\hat a$ stands for the operator of multiplication by $a$, \ie
$\hat a b=ab$, and $\lbr \alpha \beta$ stands for the (super)commutator of
$\alpha$ and $\beta$, \ie $\lbr\alpha\beta=
\alpha\beta-(-1)^{\alpha\beta}\beta\alpha$. If $\alpha$ is an even (\ie
parity preserving) operator, it follows from (\ref{sbr}) that
\begin{equation}
\alpha(ab)=\alpha(a)b+a\alpha(b).\nonumber
\end{equation}
If $\alpha$ is an odd (parity reversing) operator, then
\begin{equation}
\alpha(ab)=\alpha(a)b+(-1)^aa\alpha(b).\nonumber
\end{equation}
An operator $\alpha$ acting on $\A$ will be called a second order
(super)derivation if the expression
\begin{equation}
\lbr\alpha{\hat a}b-\alpha(a)b\nonumber
\end{equation}
is a (super)derivation on $\A$ for fixed $a$. It follows immediately
from (\ref{wb}) and (\ref{ld}) that the operator $\Delta$ in a
BV-algebra is a second order derivation. This remark leads to a
description of BV-algebras, which is given by the following theorem:
\begin{thm} Suppose that $\A$ is a super commutative,
associative  algebra and that
an odd operator $\Delta$ on $\A$ is an second order derivation and obeys
$\Delta^2=0$. Define the bracket $\{f,g\}$ by the formula
\begin{equation}
\{f,g\}=(-1)^f\Delta(fg)+(-1)^{f+1}(\Delta f)g-f\Delta g\label{swiii1}
\end{equation}
Then the operation $\{f,g\}$ satisfies (\ref{ga}) and (\ref{ji}) (
super anticommutativity and the super Jacobi identity), so that $\A$,
equipped with this bracket, is a BV-algebra.
\end{thm}
\begin{pf}
To show equation~(\ref{ga}), we use the super commutativity of \A\ and
equation~(\ref{wb}) to yield:
\begin{eqnarray*}
(-1)^g\{g,f\}&=&\Delta(gf)-(\Delta g)f +(-1)^{g+1}g\Delta f\\
&=&(-1)^{gf}\Delta(fg)+(-1)^{(g+1)f+1}f\Delta g +(-1)^{fg+1}(\Delta f)g\\
&=&(-1)^{gf}(\Delta f)g +(-1)^{(g+1)f}f\Delta g + (-1)^{(g+1)f}\{f,g\}\\
&&+(-1)^{(g+1)f+1}f\Delta g +(-1)^{fg+1}(\Delta f)g\\
&=&(-1)^{(g+1)f}\{f,g\}
\end{eqnarray*}
from which the desired equality follows immediately.
To facilitate the remainder of the proof, we first prove a lemma
\begin{lma}
The operator $\Delta$ is a super derivation operator on its bracket,\ie
\begin{equation}
\Delta\{f,g\}=\{\Delta f,g\} +(-1)^{f+1}\{f,\Delta g\}
\end{equation}
\end{lma}
\begin{pf}
This depends only on the defining property in
equation~(\ref{wb}) of the bracket
and the condition $\Delta^2=0$. To wit:
\begin{eqnarray*}
0&=&\Delta^2(fg)=\Delta[(\Delta f)g +(-1)^ff\Delta g +(-1)^f\{f,g\}]\\
&=&(\Delta^2 f)g +(-1)^{f+1}(\Delta f)\Delta g +(-1)^{f+1}\{\Delta f,g\}\\
&&+(-1)^f(\Delta f)\Delta g+f\Delta^2 g+\{f,\Delta g\}+(-1)^f\Delta\{f,g\}\\
&=&(-1)^{f+1}\{\Delta f,g\}+\{f,\Delta g\} +(-1)^f\Delta\{f,g\}
\end{eqnarray*}
\end{pf}
The proof of equation~(\ref{ji}) follows from the above lemma and the
following:
{\arraycolsep=1pt
\begin{eqnarray}
(-1)^g\{f,\{g,h\}\}&=&\{f,(-1)^g\{g,h\}\}\nonumber\\
&=&\{f,\Delta(gh)-(\Delta g)h +
(-1)^{g+1}g\Delta h\}\nonumber\\
&=&\{f,\Delta(gh)\}-\{f,(\Delta g)h\} +(-1)^{g+1}\{f,g\Delta h\}\\
\Delta\{f,gh\}&=&\{\Delta f,g\}+(-1)^{f+1}\{f,\Delta(gh)\}\\
\{f,\Delta(gh)\}&=&(-1)^{f+1}\Delta\{f,gh\}+(-1)^f\{\Delta f,gh\}\\
(-1)^g\{f,\{g,h\}\}&=&(-1)^{f+1}\Delta\{f,gh\}
+(-1)^f\{\Delta f,gh\}\nonumber\\
&&-\{f,(\Delta g)h\}+(-1)^{g+1}\{f,g\Delta h\}\\
\Delta\{f,gh\}&=&\Delta[\{f,g\}h+(-1)^{(f+1)g}g\{f,h\}\nonumber\\
&=&(\Delta\{f,g\})h +(-1)^{f+g+1}\{f,g\}\Delta h \nonumber\\
&&+(-1)^{f+g+1}\{\{f,g\}h\}+(-1)^{(f+1)g}(\Delta g)\{f,h\} \nonumber\\
&&+(-1)^{fg}g\Delta\{f,h\} +(-1)^{fg}\{g\{f,h\}\}\nonumber\\
&=&\{\Delta f,g\}h +(-1)^{f+1}\{f,\Delta g\}h \nonumber \\
&&+(-1)^{f+g+1}\{f,g\}\Delta h+(-1)^{f+g+1}\{\{f,g\}h\} \nonumber \\
&&+(-1)^{(f+1)g}(\Delta g)\{f,h\}+(-1)^{fg}g\{\Delta f,h\} \nonumber\\
&&+(-1)^{f(g+1)+1}g\{f,\Delta h\}+(-1)^{fg}\{g\{f,h\}\}\\
\lefteqn{(-1)^f\{\Delta f,gh\}
-\{f,(\Delta g)h\}+(-1)^{g+1}\{f,g\Delta h\}=}\nonumber\\
&&(-1)^f\{\Delta f,gh\}+(-1)^{f(g+1)}g\{\Delta f,h\}
-\{f,\Delta g\}h\nonumber\\
&&+(-1)^{(f+1)(g+1)+1}\Delta g)\{f,h\}+(-1)^{g+1}\{f,g\}\Delta h\nonumber\\
&&+(-1)^{fg+1}g\{f,\Delta h\}\\
(-1)^{f+1}\Delta\{f,gh\}&=&(-1)^{f+1}\{\Delta f,g\}h +\{f,\Delta g\}h
+(-1)^g\{f,g\}\Delta h+\nonumber\\
&&(-1)^g\{\{f,g\}h\}+(-1)^{(f+1)(g+1)}(\Delta g)\{f,h\}\nonumber\\
+(-1)^{f(g+1)+1}g\{\Delta f,h\}
&&+(-1)^{fg}g\{f,\Delta h\}  \nonumber\\
&&+(-1)^{f(g+1)+1}\{g,\{f,h\}\}\\
(-1)^g\{f\{g,h\}\}&\!=\!&(-1)^g\{\{f,g\}h\} \!+
\!(-1)^g(-1)^{(f+1)(g+1)}\{g\{f,h\}\}\\
\{f\{g,h\}\}&=&\{\{f,g\}h\} +(-1)^{(f+1)(g+1)}\{g\{f,h\}\}
\end{eqnarray}
}
\end{pf}
Note that the proof of equation~(\ref{ji}) depends only on the lemma and
the derivation property given in equation~(\ref{ld}), while
the lemma does not depend on super anticommutativity of the bracket. The
proof of the super anticommutativity of the bracket is the only place where
the fact that \A\ is super commutative, not merely a graded algebra,
is used.
Therefore, we see that for a general associative,
$\Z_2$-graded algebra \A, if the operator
$\Delta$ is defined as above, satisfying equations (\ref{wb}) and (\ref{ld})
as well as $\Delta^2=0$,
then the super Jacobi identity
equation (\ref{ji}) holds. Thus we may in this case extend the construction
to the case of arbitrary graded algebras. Of course, the form of the
super Jacobi identity used here will not in
general be equivalent to other standard forms
unless super anticommutativity of the bracket holds. However, it is clear
that super anticommutativity of the bracket is sufficient for both conditions
of the theorem to hold. In this case we note that the bracket also
satisfies the right derivation rule
$ \{fg,h\}=(-1)^{(h+1)g}\{f,h\}g+f\{g,h\} $.
This is a simple consequence of (\ref{ld}) and (\ref{ga}). All versions
of the super Jacobi identity will also hold, since they derive from the
super anticommutativity and any version of the super Jacobi identity.

Now we can turn to the proof of the results of \cite{zuk}.
Let us begin with a chiral algebra $V$. Recall that the structure of a chiral
algebra on a linear superspace $V$ is determined by a linear map which
assigns to each element $\ph\in V$ a formal power series
$\ph(z)=\sum \ph_{(n)}z^{-n-1}$,
where the $\ph_{(n)}$ are linear operators on $V$ and
$\ph_{(0)}=\ph$. The elements of $V$ are interpreted as states, the series
$\ph(z)$ as a field. It is assumed that there is an operator expansion (OPE) of
the product of two fields, and that OPE satisfies the associativity
condition. It is also assumed that there
are even elements 1 and $L$ in $V$ such that the corresponding fields
$1(z)$ and $L(z)$ can be considered as the unit operator and the
energy-momentum
tensor, in other words, these operators satisfy the appropriate OPE
conditions.

We consider the case when $V$ is $\Z$ graded, $V=\sum V^k$. We say that an
element $\ph\in V^k$ and its corresponding field $\ph(z)$ have the weight
(conformal dimension) $k$, denoting the weight of $\ph$ by $\Delta_\ph$.
We assume that for
$\ph \in V^k$ the operator $\ph_{(n)}$ acts from $V^r$ to $V^{r-n+k}$. The
weight of $1$ is equal to zero, the weight of $L$ is equal to 2. If $\ph$
is a field of weight $k$, it is convenient to introduce the notation
$\ph_n=\ph_{(n+k-1)}$. Then the field $\ph$ can be represented in the
form $\ph=\sum\ph_nz^{-n-k}$.
We do not give
a detailed definition of the chiral algebra, referring to  \cite{zuk} or
\cite{flm}.

Let us write down, however, some relations that will be used later:
\begin{eqnarray}
L(z)&=&\sum L_nz^{-n-2}
\label{swii1}
\end{eqnarray}
where the $L_n$ are the generators of the Virasoro algebra.
\begin{equation}
(L_0\psi)(z)=\Delta_\psi\psi(z),\quad
\quad (L_{-1}\psi)(z)=\frac{d\psi(z)}{dz}\label{new}
\end{equation}
\begin{eqnarray}
\psi_{(1)}(1)&=&\psi_{\Delta_\psi}(1)=1\nonumber\\
\lbr{\psi_{(m)}}{\ph_{(n)}}&=&
\sum_{i\in\N}\pmatrix{m\cr i}(\psi_{(i)}\ph)_{({m+n-i})}\label{sw1}
\end{eqnarray}
In particular,
\begin{eqnarray}
\lbr{\psi_{(0)}}{\ph_{(n)}}&=&(\psi_{(0)},\ph)_{(n)}\label{sw2}\\
\lbr{\psi_{(1)}}{\ph_{(n)}}&=&
(\psi_{(1)}\ph)_{(n)}+(\psi_{(0)}\ph)_{({n+1})} \label{sw3}
\end{eqnarray}
One can derive (\ref{sw1}) from the "Jacobi identity" of \cite{zuk}, by
taking $f(z,w)=z^mw^n$ (see \cite{flm}).
It follows from (\ref{sw2}) that
\begin{equation}
\lbr{\psi_{(0)}}{\ph(z)}\xi=(\psi_{(0)}\ph)(z)\xi\label{swi1}
\end{equation}
In other words, $\psi_{(0)}$ acts as a (super) derivation on the product
$\psi(z)\xi$. Analogously, we get from (\ref{sw3}) that $\psi_{(1)}$ acts as
a second order derivation of $\psi(z)\ph$. More precisely,
\begin{equation}
\lbr{\psi_{(1)}}{\ph(z)}\xi-(\psi_{(1)}\ph)(z)(\xi)=z(\psi_{(0)}\ph)(z)\xi
\label{swi2}
\end{equation}

Let us suppose now that an odd operator $Q$ satisfying $Q^2=0$, acts on the
chiral algebra $V$, and that $Q$ satisifies the condition
\begin{equation}
\lbr Q{\psi(z)}\ph=(Q\psi)(z)\ph\label{sw4}
\end{equation}
Usually $Q$ can be represented in the form
$Q=\oint \sigma(z)dz=\sigma_0$,
where $\sigma(z)$ is an odd field of weight 1; then (\ref{sw4}) follows
from (\ref{swi1}).
Suppose  that there is an odd field $b$ of weight 2 obeying
\begin{eqnarray}
L(z)=\lbr{Q}{b(z)}\label{sw5}\\
\lbr{b(z)}{b(z')}=0\nonumber
\end{eqnarray}
Hence $\lbr LQ =0$ and $Q$ preserves the weight.
We say then that $V$ is a topological chiral algebra.

It is well known that one can obtain a topological chiral algebra from
any chiral algebra with central charge 26 by adding ghosts,
 and from any $N=2$ superconformal field
theory.

It is proved in \cite{zuk} that the superspace $H=\ker Q/Im Q$ (the homology
of $Q$), can be equipped with the structure of a BV-algebra.
We will give a simple proof of this result. Let us first note
that, as follows from (\ref{sw4}), for $\psi,\ph \in\ker Q$,  the product
$\psi(z)\ph$ satisfies $Q(\psi(z)\ph)=0$, and therefore generates an
element $\psi\circ\ph$ of the space $H\lbr z{z^{-1}}$ of formal Laurent
series with coefficients in $H$. This element is unchanged if we replace
$\psi$ and $\ph$ by $\psi+Q\alpha$ and $\ph+Q\beta$ (again, this follows
from (\ref{sw4})).
Using (\ref{sw5}) and (\ref{new}), we obtain
\begin{equation}
\frac d{dz}(\psi(z)\ph)=
(L_{-1}\psi)(z)\ph=
(\lbr Q{b_{-1}}\psi)(z)\ph=
(Qb_{-1}\psi)(z)\ph=
Q((b_{-1}\psi)(z)\ph)
\label{swii2}
\end{equation}
It follows from (\ref{swii2}) that $\frac d{dz}(\psi\circ\ph)=0$, and therefore
$\psi\circ\ph$ can be considered as an element of $H$. Thus we can define
a product $a\circ b$ of elements $a,b\in H$, as the homology class
of $\psi\circ\ph$, where $\psi$ and $\ph$ are representatives of the
homology classes of $a$ and $b$, resp.
It is easy to verify that the product $a\circ b$ coincides with the dot product
$a\cdot b$ defined in \cite{zuk}, as the homology class of
\begin{equation}
\oint\psi(z)\ph\frac{dz}{2\pi iz}=\int^1_0\psi(e^{2\pi it})\ph dt
\label{swii3}
\end{equation}
where $\psi\in a$ and $\ph \in b$. To prove this statement, we note that
$\int^1_0\psi(ze^{2\pi it})\ph dt$ is homologous to $\psi(z)\ph$.
This is a particular case of a more general assertion:
if $Q\alpha(t)=0$ for $0\le t\le1$ and $\alpha'(t)=Q\beta(t)$, then
$\int_0^1 \alpha(t)  dt-\alpha(0)=Q\int^1_0\beta(t)  dt$,
and therefore $\int_0^1\alpha(t) dt$ is homologous to $\alpha(0)$.
(Of course, we have to assume that $\beta(t)$ is an integrable function
of $t$.)  To apply this assertion to the case at hand we take
$\alpha(t)=\psi(ze^{2\pi it})\ph$; it follows from (\ref{swii2})
that then one can
take $\beta(t)=(b_{-1}\psi)(ze^{2\pi it})\ph$.
In this manner we have shown that the homology class of
$\int^1_0\psi(ze^{2\pi it})\ph dt$
coincides with the homology class of
$\psi(z)\ph$, in particular, it does not depend on $z$.
We see that this homology class coincides with
(\ref{swii3}), because (\ref{swii3}) can be obtained by
means of substitution of
$z=1$ into
$\int^1_0\psi(ze^{2\pi it})\ph dt$.

The multiplication
$(a,b)\mapsto a\circ b$ in $H$ is both associative and distributive.
The distributivity is evident, the associativity follows from the
associativity of OPE. To be more exact, OPE is not associative in the
standard sense; the associativity is expressed by the equation
\begin{equation}
(u(z_1)v)(z_2)w=u(z_1+z_2)(v(z_2)w)\label{swii2a}
\end{equation}
If $u$, $v$, $w$ are representatives of the homology classes
$\ph$, $\psi$, $\zeta$ resp., then the left hand side belongs to the
homology class $(\ph\circ\psi)\circ(\zeta)$,
 and the right hand side
belongs to the homology class
$\ph\circ(\psi\circ\zeta)$.
(To check this one should use (\ref{swii2}).)
Therefore, associativity follows from (\ref{swii2a}).

To prove super commutativity of the product in $H$ we use the
equivalence of our definition of this product and the definition in
\cite{zuk},
and refer the reader to the six line proof in \cite{zuk}.
Later we will give a geometric interpretation of the
product, which will make the super commutativity almost obvious. Thus we
have provided $H$ with the structure of a super commutative, associative
algebra.

Let us note that the operator $b_0$ can be considered as an
operator acting on $H$. Reallly, using
\begin{equation}
L_0=\lbr Q{b_0}\label{new1}
\end{equation}
we obtain from $Q\psi=0$, and $L_0\psi=\lambda \psi$, with
$\lambda\ne0$, that $\psi=Q(\lambda^{-1}b_0\ph)$.
This means that each homology class $x\in H$ has a representative
$\xi\in\A$ satisfying $L_0\xi=0$.  Using (\ref{new1}) once more
we see that $Q(b_0\xi)=0$ and therefore one can define $b_0x$ as the
homology class of $b_0\xi$. It is evident that $b_0^2=0$. Using
(\ref{swi2}) we see that $b_0=b_{(1)}$ acts as a second order
derivation. Therefore we can apply the theorem above, with $b_0$ as
$\Delta$. In this manner we obtain the structure of a BV-algebra on $H$.
It follows from (\ref{swiii1}) and (\ref{swi2}) that the element
$\{x,y\}$, where $x$, $y$ $\in H$ can be written as the homology class
of
\begin{equation}
z(b_{-1}\xi)(z)\eta\label{swiii2}
\end{equation}
where $\xi$ and $\eta$ are representatives of the classes $x$ and $y$
satisfying $L_0\xi=0$ and $L_0\eta=0$. Integrating (\ref{swiii2}) over
the contour $|z|=1$, we get that (\ref{swiii2}) is homologous to
$Res_z(b_{-1}\xi)(z)(\eta)$; this form of the bracket is obtained in
\cite{zuk}.

Note that the assumption about integrality of weight was used in our
proof of the coincidence of our definition of the product in $H$ with
the definition in \cite{zuk}. (If the weight is not integral, the
integrand in  (\ref{swii3}) is multivalued.)
However, we have seen that with the calculations in $H$ we can restrict
ourselves to elements of $\A$ having zero weight. Thus we can exclude
fields with non-integral weight from our discussion.

Let us consider now a conformal algebra $\A$ with left and right Virasoro
subalgebras. This means that to every element $\ph\in\A$, we assign a
field $\ph(z,\ol z)$, considered as a formal series
\begin{equation}
\ph(z,\ol z)=\sum \ph_{(m,n)}z^{-(m+1)}{\ol z}^{-(n+1)}\label{fin}
\end{equation}
where the $\ph_{(m,n)}$ are operators acting on $\A$, and $\ph_{(0,0)}=\ph$.
The numbers $m$, $n$ in (\ref{fin}) are not necessarily integers.
We assume that for fields
$\ph(z,\ol z)$,
$\psi(z,\ol z)$
corresponding to the elements $\ph$, $\psi$ in $\A$, one can write OPE
\begin{equation}
\ph(z+\zeta,\ol z+\ol \zeta)\circ\psi(z,\ol z)=
\sum \zeta^{m}{\ol\zeta}{}^{n}C_{m,n}(z,\ol z)
\nonumber
\end{equation}
 The coefficients of OPE are given by the formula
\begin{equation}
C_{m,n}(z,\ol z)=(\ph_{(-m-1,-n-1)}\psi)(z,\ol z).
\nonumber
\end{equation}
We assume the associativity of OPE. We suppose also that there are
elements 1, $L$, $\ol L$ in $\A$ such that $L(z)$ and $\ol L(\ol z)$ are
holomorphic and anti-holomorphic parts of the energy-momentum tensor
resp., and that $1(z)$ is the unit operator. Standard OPE's for products
of these fields and other fields are assumed.  Therefore representing
$L(z)$ and $\ol L(\ol z)$ as
$L(z)=\sum L_nz^{-n-2}$ and
$\ol L(\ol z)=\sum \ol L_n\ol z^{-n-2}$, we obtain two
commuting Virasoro algebras
acting on $\A$ (with generators $L_n$ and $\ol L_n$, resp.)
We have
\begin{eqnarray}
(L_{-1}\ph)(z,\ol z)&=&\frac{\partial \ph(z,\ol z)}{\partial z}\\
(\ol L_{-1}\ph)(z,\ol z)&=&\frac{\partial \ph(z,\ol z)}{\partial \ol
z}.
\end{eqnarray}

We suppose that the space $\A$ can be represented as a direct sum of
subspaces $\A_{m,n}$ consisting of eigenvectors of the commuting operators
$L_0$, $\ol L_0$, with eigenvalues $m$ and $n$, resp; the pair
$(m,n)=(\Delta_\psi,\Delta_\ph)$ is
called a weight (or conformal dimension) of $\ph\in\A_{m,n}$. If
$\ph\in\A_{k,l}$, then the operator $\ph_{(m,n)}$ has to act from
$\A_{r,s}$ to $\A_{r-m+k,s-n+l}$.
It follows from our assumptions that
\begin{eqnarray}
(L_0\ph)(z,\ol z)=\Delta_\ph\ph(z,\ol z)\nonumber\\
({\ol L}_0\ph)(z,\ol z)={\ol\Delta}_\ph\ph(z,\ol z)\label{new3}
\end{eqnarray}
Let us emphasize that we don't assume that the eigenvalues of $L_0$ and
$\ol L_0$ are integers, however $\Delta_\psi-{\ol\Delta}_\ph$
must be integral to guarantee that
the fields are single valued.

We say that a conformal algebra $\A$ is a topological conformal algebra
if there are odd elements $\sigma$, $\ol\sigma$, $b$, $\ol b$ in $\A$
having weights
(1,0),
(0,1),
(2,0),
(0,2)
resp. and satisfying the conditions
\begin{equation}
\ol L_n\sigma=L_n\ol\sigma=\ol L_n b=L_n\ol b=0.
\nonumber
\end{equation}
(In other words, we have holomorphic fields $\sigma$, $b$ and
anti-holomorphic fields $\ol\sigma$ and $\ol b$.) We define
operators $Q$, $\ol Q$ by the formulas
$Q=\oint\sigma(z)dz=\sigma_{(0,0)}$,
$\ol Q=\oint\ol \sigma(\ol z)d\ol z=\ol \sigma_{(0,0)}$
and suppose that $Q^2=\ol Q^2=0$,
$\lbr Q{{\ol b}(z)}=0$,
$\lbr {\ol Q}{{\ol b}(z)}=0$,
$L(z)=\lbr Q{b(z)}$, and $\ol L(\ol z)=\lbr{\ol Q}{\ol b(\ol z)}$.
Standard OPE for $b$, $\ol b$, $L$ and $\ol L$ are assumed.

Let us consider the homology $H=\ker(Q+\ol Q)/{\rm Im}(Q+\ol Q)$
constructed by means of the operator $Q+\ol Q$ acting on $\A$.
One can define a multiplication in $H$ by repeating the construction
given in the chiral case.

It is possible to give a geometric interpretation of this multiplication.
Let us take as a starting point an axiomatic approach to
conformal field theory.  Consider a closed, oriented Riemann surface
$\Sigma$, \ie a complex curve, and $m$ holomorphic and $n$ anti-holomorphic
maps of a disc into this surface. In axiomatic field theory, these data
permit us to construct a map $\A^{\otimes m}\ra\A^{\otimes n}$.(Here,
$\A$ stands for the space of states.) In particular, if $\Sigma=S^2$,
$m=2$ and $n=1$, we obtain a map $\A\bigotimes\A\ra \A$, which can be
interpreted as a multiplication in $\A$. For every conformal algebra
$\A$\ one can construct a conformal field theory at least on the
surfaces of genus zero. The multiplication $(u,v)\ra u(z)v$ coincides
with the map $\A\bigotimes\A\ra \A$ for a specific choice of the maps of
the disc into $S^2$. If $\A$ is a topological chiral algebra, it is easy
to check that the multiplication $\A\bigotimes\A\ra \A$ corresponding to
an arbitrary choice of discs in $S^2$ determines a multiplication in the
homology group $H$, and this multiplication does not
depend on the choice of the discs. (The proof is similar to the proof of
independence of the homology class of $u(z)v$ on the choice of $z$.) The
geometric interpretation makes obvious
the associativity and super commutativity of this
multiplication. A slight modification of the considerations applied in
the chiral case shows that the operator $b_0+\ol{b}_0$ determines an odd
second order derivation in $H$ and that $(b_0+\ol{b}_0)^2=0$. This
means that this operator specifies the structure of a BV algebra in
$H$. Thus we have shown the following theorem:
\begin{thm}
Let $\A$ be a  topological conformal algebra. Then the corresponding
homology $H$ of $\A$ with respect to
$Q+\ol Q$ naturally has the structure of a
BV-algebra.
\end{thm}

The constructions above are closely related to the constructions of
string field theory. We considered here the operations in the homology group
$H$. One can consider the corresponding operations in $\A$, but the
properties of the operations in $H$ are valid in $\A$ only up to
homotopy. The consideration of higher homotopies in $\A$ should lead
to interesting algebraic structures, including the structure of
a homotopy Lie algebra, as was constructed in \cite{zwi} (see also
\cite{sta}).
However the most interesting structures are connected with the relative
homology
$H_k=\ker(Q+\ol Q)\cap\A_k/(Q+\ol Q)A_k$ where $\A_k$ denotes the subspace
of $\A$
consisting of all elements $\ph\in \A$ satisfying
\begin{eqnarray}
(L_k-\ol L_k)\ph&=&0\label{swii4}\\
(b_k-\ol b_k)\ph&=&0\nonumber
\end{eqnarray}
It is well known that one can construct string amplitudes considering
$\A$ as a conformal background;(see \cite{wit2}).
(More precisely, to construct string amplitudes one has to extend $\A$
to a topological conformal field theory, defined on surfaces of
arbitrary genus.)
Then $H_0$ is closely related to
the space of physical states.
Note that we don't impose any conditions on the ghost number of states
in $H_0$. (Moreover, we don't need to introduce the notion of ghost
number.)
It would be interesting to prove that $H_0$ can be
equipped with the structure of a BV-algebra.
This can be done in certain circumstances (see \cite{ver},
\cite{zwi}), but we don't know a general proof. However, slight
modifications of the considerations above can be used to introduce a
BV-algebra structure in $H_{-1}$.

We are indebted to M.~Kontsevich, J.~Stasheff, E.~Verlinde,
G.~Zuckerman, and B.~Zwiebach for useful discussions.


\begin{thebibliography}{99}

\bibitem{bv} Batalin, I. \& Vilkovisky, G:{\sl Gauge Algebra and
Quantization}. Physics Letters, 102B, {\bf 27}(1981).

\bibitem{flm}Frenkel, I., Lepowsky, J., Meurmann, A.:{\sl
Vertex Operator Algebras and the Monster}.
Boston: Academic Press (1988)

\bibitem{ger} Gerstenhaber, M.:{\sl The Cohomology Structure of an
Associative Ring}. Annals of Mathematics (2), {\bf 78}(1963).

\bibitem{zuk}Lian, B., Zuckerman, G.:
{\sl New Perspectives on the BRST-Algebraic
Structure of String Theory}. Preprint.

\bibitem{sc2}Schwarz, A:{\sl The Geometry of Batalin-Vilkovisky
Quantization}. Commum. Math. Phys. (To appear).

\bibitem{sch}Schwarz, A.:{\sl Semiclassical Approximation in
Batalin-Vilkovisky Formalism}. Preprint.

\bibitem{sta}Stasheff, J.:{\sl Towards a Closed String Field Theory:
Topology and Convolution Algebra}. UNC-MATH 90/1.

\bibitem{ver}Verlinde, E.:{\sl The Master Equation of 2D String Theory}.
Preprint. IASSNS-HEP-9214.

\bibitem{wit}Witten, E.:{\sl A Note on the Antibracket Formalism}.
Mod. Phys. Lett. A5,{\bf 487}(1990).

\bibitem{wit2}Witten, E.:{\sl Chern-Simons Gauge Theory as a String
Theory}. Preprint. IASSNS-HEP-92145.

\bibitem{zwi}Zwiebach, B.:{\sl Closed String Field Theory: Quantum Action
and the BV Master Equation}. Preprint. IASSNS-HEP-92141.

\end{thebibliography}
\end{document}